\documentclass[]{spie}  

\usepackage{amsmath,amsfonts,amssymb}
\usepackage{graphicx}
\usepackage[colorlinks=true, allcolors=blue]{hyperref}
\usepackage{adjustbox}
\usepackage{indentfirst}

\usepackage{xcolor}

\usepackage{etoolbox}
\usepackage{multirow}
\usepackage[normalem]{ulem}
\useunder{\uline}{\ul}{}
\usepackage{threeparttable}

\apptocmd{\thebibliography}{\setlength{\itemsep}{10pt}}{}{}

\title{ESFPNet: efficient deep learning architecture for real-time lesion segmentation in autofluorescence bronchoscopic video}

\author{Qi Chang,$^a$ Danish Ahmad,$^b$ Jennifer Toth,$^b$
\vspace*{-10pt}Rebecca Bascom,$^b$ and William E. Higgins$^{a,*}$\\
\vspace*{8pt}$^a$School of Electrical Engineering and Computer Science\\
$^b$College of Medicine\\
Penn State University, University Park and Hershey, PA
}

\authorinfo{$^*$Correspondence: Email: weh2@psu.edu;
	WWW: http://mipl.ee.psu.edu/; Telephone: 814-865-0186}

\begin{document}

\maketitle

\begin{abstract}
\noindent Lung cancer tends to be detected at an advanced stage, resulting in a high
patient mortality rate.  Thus, much recent research has focused on early disease detection.
Lung cancer generally first appears as lesions developing within the bronchial
epithelium of the airway walls.  Bronchoscopy is the procedure of choice for
effective noninvasive bronchial lesion detection.  In particular,
autofluorescence bronchoscopy (AFB) discriminates the autofluorescence properties of
normal and diseased tissue, whereby lesions appear reddish brown in AFB video frames,
while normal tissue appears green.  Because recent studies show AFB's high sensitivity
for lesion detection, it has become a potentially pivotal method during the standard
bronchoscopic airway exam for early-stage lung cancer detection.
Unfortunately, manual inspection of AFB video is extremely tedious and error prone,
while limited effort has been expended toward potentially more robust automatic AFB
lesion detection and segmentation.
We propose a real-time deep-learning architecture dubbed ESFPNet for accurate segmentation and robust detection of bronchial lesions in an AFB video stream.
The architecture features an encoder structure that exploits pretrained Mix Transformer (MiT) encoders and an efficient stage-wise feature pyramid (ESFP) decoder structure.
Segmentation results derived from the AFB airway-exam videos of 20 lung cancer patients
indicate that our approach gives a mean Dice index = 0.756 and average Intersection of
Union = 0.624, results that are superior to those generated by other recent architectures.
In addition, our method enables a processing throughput of
27 frames/sec.  Thus, ESFPNet gives the physician a potential tool for confident
real-time lesion segmentation and detection during a live bronchoscopic airway exam.
Moreover, our model shows promising potential applicability to other domains, as evidenced
by its state-of-the-art (SOTA) performance on the CVC-ClinicDB and ETIS-LaribPolypDB datasets
and its superior performance on the Kvasir and CVC-ColonDB datasets.
\end{abstract}

\keywords{bronchoscopy, lung cancer, lesion segmentation and detection, airway wall analysis, autofluorescence imaging, deep learning, mix transformer, efficient stage-wise
feature pyramid}


\section{INTRODUCTION}
\label{sec:intro}
Lung cancer is the most common cause of cancer death worldwide \cite{Bray-CA2018}.
An important goal toward improving lung cancer survival is to detect the disease at an early
stage, thereby giving an opportunity for the most effective treatment options.
Lung cancer begins when lesions develop in the bronchial epithelium of the lung mucosa.
These bronchial lesions, which can eventually evolve into squamous cell lung cancer, also
help predict the potential development of other lung cancers.  Hence, methods for
early detection of bronchial lesions are essential to help improve lung cancer patient care.
A noninvasive way for physicians to search for such lesions is to use bronchoscopy for
imaging the airway epithelium during a routine airway exam \cite{Inage-CCM2018}.  Among
the current advanced videobronchosopic techniques, autofluorescence bronchoscopy (AFB)
exhibits high sensitivity to suspicious bronchial lesions and effectively distinguishes developing bronchial lesions from the normal epithelium.  In AFB video, lesions appear
reddish brown, while normal tissue appears green. Unfortunately, current standard practice
entails manual inspection of an incoming AFB video stream, which is extremely tedious
and error prone.

A promising strategy for improving this situation is to consider
computer-based lesion analysis methods for AFB video frames.  A simple, but
largely ineffective, approach applied in some clinical work is to measure the red/green
ratio \cite{Kusunoki-Chest2000}.  Other more recent work has considered standard
computer-based image processing methods and/or conventional machine learning techniques
\cite{Finkvst-SVJME2012,Finkst-JME2017,Chang-EMBC2020}:
Unfortunately, these works all have at least one of the following limitations:
\begin{enumerate}
\item Need complicated image preprocessing, including image enhancement and/or feature
extraction before reaching lesion decisions.
\item Do not provide robust, accurate segmentation of abnormal lesion regions as an aid toward locating potential lesions.
\item Can't process an input AFB video stream in real-time, thereby making the methods unsuitable for making lesion decisions during a live bronchoscopic airway exam.
\end{enumerate}
These limitations make the methods unsuitable for lesion segmentation and detection during
a live bronchoscopic airway exam.

Recent deep-learning-based architectures, which have achieved great success in medical
image analysis, show much promise for handling these limitations.  As an example,
Unet++ is a powerful and widely used architecture for semantic medical image
segmentation \cite{Zhou-DL2018}.
The Unet++ adds efficient and densely-connected nested-decoder subnetworks to the
popular Unet architecture.\cite{Ronneberger-MICCAI2015}
It also applies a deep supervision mechanism that allows
for improved aggregation of features across different semantic scales.
Although Unet++ can provide more accurate segmentations than Unet, the requisite dense
connections demand extensive computation.

As another example, the Caranet also utilizes deep supervision to
enhance the use of aggregative features \cite{Lou-arXiv2021cara}.  Yet, in
contrast to the complicated sub-networks of Unet++, the Caranet includes the
advantageous self-attention mechanism and draws on the context axial reverse attention technique on a pretrained Res2Net backbone.  Hence, it has been shown to enable faster
processing time and better segmentation performance than Unet++ when tested over
multiple public medical datasets.  Nevertheless, the design of the self-attention
mechanism in the Caranet is complex.  On another front, the Segformer has shown much
success in the semantic segmentation domain \cite{Xie-NeurIPS2021}.  The Segformer
provides a simple and efficient layout that utilizes the attention technique referred
to as ``Mix Transformer (MiT) encoders."  In a later development, the SSFormer
architecture extracts aggregate local and global step-wise features from pretrained
MiT encoders to predict abnormal regions \cite{Wang-arXiv2022}.  Tests with
the CVC-ColonDB and Kvasir-SEG datasets demonstrate the generalizability and superior
performance of the SSFormer (and its use of the MiT encoders) over Caranet for
medical imaging applications.  Yet, the feature pyramid used by SSFormer could be made
more efficient, thereby reducing the processing time and network complexity.
Note that the number of parameters defining a network gives a direct indication of
the number of floating-point operations (FLOPS) needed to process an input and, hence,
its computational efficiency (execution time).  (Table \ref{table:Comparison} later
illustrates this point.)

We propose a more efficient deep-learning-based architecture that enables
real-time segmentation and detection of bronchial lesions in an AFB video.
Our architecture draws on pretrained Mix Transformer (MiT) encoders as the backbone
and a decoder structure that incorporates an efficient stage-wise feature pyramid
(ESFP) to help promote accurate lesion segmentation.

\section{METHODS}
\label{sec:method}

For a given AFB video sequence, our goal is achieve both high lesion detection accuracy
and high computational throughput.  At the outset, we point out that state-of-the-art
deep-learning-based neural network architectures generally require a large amount
of data to adequately train and test them (the ``data hunger" problem).  This is
because of the large number of network parameters that require precise tuning during
the training process. Unfortunately, for our AFB video analysis application, while we
seemingly have a large dataset available to training and testing, the amount of data we
have still proves to be insufficient.  Our proposed approach offers a solution to
all of these issues.

\begin{figure}[htb]
	\centering
	\includegraphics[width=0.95\textwidth]{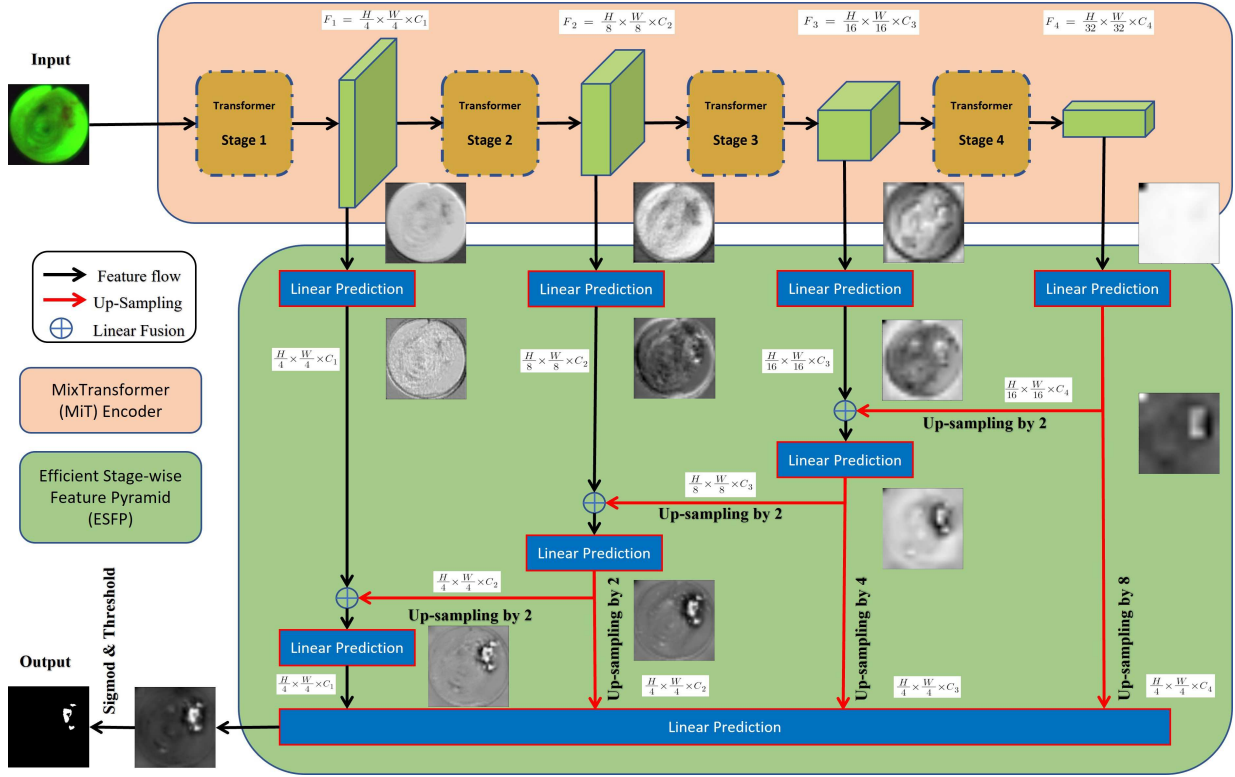}%
    \vspace*{6pt}
	\caption{Block diagram of the ESPFNet architecture.}
	\label{fig:network}
\end{figure}

Figure \ref{fig:network} depicts the proposed ESFPNet architecture, which utilizes the
Mix Transformer (MiT) encoder as the backbone and uses an efficient stage-wise feature pyramid (ESFP) as the decoder to generate segmentation outputs.  The basic input is
a raw AFB video frame, while the output is a frame that either presents a segmented
detected lesion or no output (a normal frame).

Sections \ref{sec:method:backbone}-\ref{sec:method:ESFP} describe each component
of the architecture.  Section \ref{sec:method:imple} then discusses implementation
details.

\subsection{Backbone}
\label{sec:method:backbone}

CNN-based encoders, such as the Unet and SegNet implementations, have enjoyed much success for image segmentation tasks (CNN = convolutional neural network)
\cite{Ronneberger-MICCAI2015,Badrinarayanan-TPAMI2017}.
A CNN-based encoder, motivated by the idea that every image pixel depends on its
neighboring pixels, uses filters on an image patch to extract relevant local features.
Yet, if a processing model utilized all image data instead of only the patches considered
by the filters, then  processing performance would be expected to improve.  This
concept helps the so-called Vision Transformers (ViT) work better than most CNN
models \cite{Yuan-ICCV2021}.  The Mix Transformer encoder (MiT) is a module that
takes advantage of the idea of the ViT network and uses four overlapping path-merging
modules and self-attention prediction in four stages \cite{Xie-NeurIPS2021}.
These stages not only furnish high-resolution coarse features, but also provide
low-resolution fine-grained features.  In addition, the high-  and low-resolution
features are commonly used to boost the performance of semantic segmentation.

On the other hand, the limitation of using transformers as encoders is obvious.
The self-attention layers used by transformers lack locality inductive bias (the notion
that image pixels are locally correlated and that their correlation maps are translation-invariant) and lead to the problem of data hunger \cite{Yuan-ICCV2021}.
To alleviate the challenge of data hunger for applications limited by small datasets,
one can exploit the widely used concept of transfer learning.  The MiT encoders, which
take advantage of this idea, are pretrained on the large ImageNet database
\cite{Deng-CVPR2009}.  For our ESPFNet architecture, we integrate these pretrained
MiT encoders as our backbone and train them again with the initialized decoders.
This proves to be a straightforward way to facilitate good performance over our
small task-specific datasets, while also being able to exceed the performance of
state-of-the-art CNN models.

\subsection{Efficient stage-wise feature pyramid (ESFP)}
\label{sec:method:ESFP}

The prediction results of the decoder rely on multi-level features from the encoder,
where local (low-level) features are extracted from the shallow parts of the
encoders, while global (high-level) features are extracted from the deeper parts.
Previous research has demonstrated that the sufficiency of local features obtained
in the shallow part of the transformer directly affects the model's performance
\cite{Raghu-NeurIPS2021}.  The existing Segformer model, however, equally concatenates
these multi-level features to predict segmentation results and, hence, lacks the
ability to sufficiently and selectively use the local features \cite{Xie-NeurIPS2021}.
To address this issue, the SSFormer architecture includes an aggregating feature
pyramid architecture that first utilizes two layers of convolution to preprocess
feature outputs from each MiT stage.  It then fuses any two features in reverse order
(from deep to shallow) until final prediction \cite{Wang-arXiv2022}.  In this way,
local features gradually guide the model's attention to critical regions.

We point out, however, that high-level (global) features contribute more to overall
segmentation performance than low-level (local) features.  Although SSFormer enhances
the contribution of local features, the usage of high-level features is weakened.
Furthermore, the SSFormer architecture's usage of the two convolution layers is
inefficient.  Inspired by the structure of a lightweight channel-wise feature pyramid network (CfpNet) that fuses every pair of features and concatenates multi-level fused elements for the final prediction, we propose a novel and efficient stage-wise feature pyramid (ESFP)
to exploit multi-stage features \cite{Lou-arXiv2021cfp}.
ESFP starts with linear predictions of each stage output (efficient in the number of connecting channels) and then linearly fuses these preprocessed features from global
to local.  These intermediate aggregating features are concatenated and work
cooperatively to produce final segmentations.

As a final comment, we tested several different versions of the proposed ESFPNet
architecture based of the different MiT encoder scales available \cite{Xie-NeurIPS2021}:
ESFPNet-T (tiny model), ESFPNet-S (standard model), and ESFPNet-L (large model),
which are based on the MiT-B0, -B2, and -B4 encoders, respectively.

\subsection{Implementation details}
\label{sec:method:imple}

We implemented our model in Pytorch and accelerated training via NVIDIA GPUs.
Given the difference in MiT encoders' scales and the GPU memory required to train, we
trained these networks either on an NVIDIA RTX 3090 or on an NVIDIA TESLA A100 GPU.
Before training, we resized the inputs to $352\times 352$ pixels and normalized them
for segmentation.  We also employed random flipping, rotation, and brightness changing
as data augmentation operations on the inputs.
Our loss function combines the weighted intersection over union (IoU) loss and the weighted binary cross-entropy (BCE) loss:
\begin{equation}
L = L^{w}_{\rm IoU} + L^{w}_{\rm BCE} \ .
\end{equation}
We used the default AdamW optimizer with the learning rate $1e^{-4}$ and trained our models
for 200 epochs.

Details for the training and validation data sets appear in
Section \ref{sec:val}.  Next,  Section \ref{sec:results} gives a series of experiments
that not only measure the performance of our proposed ESFPNET model, but also compares
it to other recent deep learning models.

\section{Experimental Datasets and Metrics}
\label{sec:val}

This section summarizes the test datasets and validation metrics used for the experiments.
Section \ref{subsec:AFBdata} summarizes our AFB dataset.  As an
additional validation experiment to measure our network's general applicability
to other domains, we also performed tests with five public datasets, as
summarized in Section \ref{subsec:Polypdata}: Kvasir\cite{Jha-MMM2019},
CVC-ClinicDB\cite{Bernal-CMIG2015}, CVC-T\cite{Vazquez-JHE2017}, CVC-ColonDB\cite{Tajbakhsh-TMI2015}, and
ETIS-LaribPolypDB\cite{Silva-IJCARS2014}.  Finally, Section \ref{subsec:Metrics}
discusses the test metrics employed.

\subsection{AFB image dataset} \label{subsec:AFBdata}

Table \ref{table:AFBDataset} summarizes the details of our AFB image dataset.  For
our dataset,  we isolated 208 lesion frames (frame size, $720\times720$), which
depicted clear ground truth bronchial lesions, and also selected 477 normal frames.
These frames were drawn from the AFB airway exams from 20 lung cancer patients seen at
our University Hospital.  These cases were collected under informed consent and an
approved IRB protocol.  Because a lesion typically appears in many frames within a
given video sequence, we selected Lesion frames to ensure variations in airway
location, size, and viewing direction.  Also, all frames were labeled by an expert.
Regarding the normal frames, we selected these randomly around the vicinity of each
observed airway bifurcation, taking care not involve any lesions or lesion
components in selected frames.

As Table \ref{table:AFBDataset} shows, we split the AFB dataset into train,
test, and validation subsets using approximately a 50\%, 25\%, and 25\% split,
respectively.  Every lesion and normal frame from a given case was placed in the
same subset to guarantee independence between training, testing, and validation phases.
Lastly, the lesion sizes summarized in Table \ref{table:AFBDataset} are given as
percentages, where 100\% = 352$\times$352$\times \pi$ pixels (this is the pixel-based
area of an AFB frame's circular scan region).

\begin{table}[!h]
\begin{center}
\scalebox{0.78}{%
\begin{tabular}{c||c|c|c|c}
Dataset    & case / (lesion frames, normal frames)                                                                                                                                                        & total lesion / normal frames & lesion/normal split ratio & lesion size range \\ \hline \hline
Train      & \begin{tabular}[c]{@{}c@{}}156 / (35,16); 176 / (8,24); 178 / (7,17);\\ 192 / (19,24); 188 / (12,11); 172 / (0,39);\\ 189 / (0,30); 191 / (0,26); 182 / (5,22);\\ 187 / (11,14)\end{tabular} & 97 / 223                     & 46.7\% / 46.8\%                    & 0.31\% - 54.30\%  \\ \hline
Validation & \begin{tabular}[c]{@{}c@{}}171 / (58,3); 181 / (0,43); 173 / (0,35);\\ 179 / (0,35); 180 / (0,23)\end{tabular}                                                                               & 58 / 139                     & 27.9\% / 27.1\%                    & 0.20\% - 75.15\%  \\ \hline
Test       & \begin{tabular}[c]{@{}c@{}}184 / (33,3); 174 / (9,25); 195 / (11,16);\\ 157 / (0,35); 190 / (0,36)\end{tabular}                                                                              & 53 / 115                     & 25.4\% / 24.1\%                    & 0.47\% - 45.87\%  \\
\end{tabular}}
\end{center}
\caption{\baselineskip=8pt AFB dataset details.  ``case" = patient case number, ``lesion/normal frames" = the number of lesion or normal frames selected from a case,
``total lesion/normal frames" = the total number of selected lesion or normal frames
for a particular data subset, while ``lesion/normal split ratio" gives the percentage
of frames in a given class (lesion or normal) assigned to a particular data subset.
Finally, ``lesion size range" indicates the range of lesion sizes encountered in a
particular subset, where the values indicate what percent of a frame's circular
scan area contains a lesion.}
\label{table:AFBDataset}
\end{table}

\begin{figure} [!htb]
	\vspace*{-2pt}
	\begin{center}
		\begin{tabular}{c}
		\includegraphics[width=6.5in]{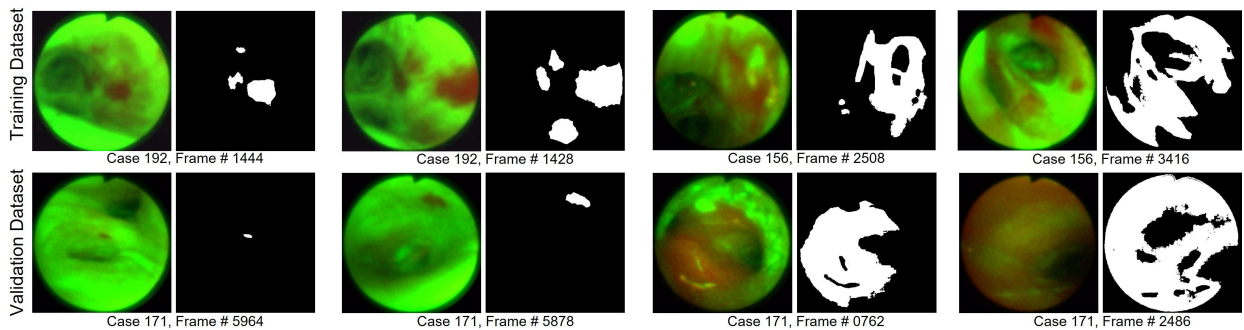} \\
		\end{tabular}
	\end{center}
	\vspace*{6pt}
	\caption[]{\baselineskip=8pt AFB lesion examples from four cases:  two from
the training subset (first row) and two from the validation subset (second row).
Each frame pair gives the original frame (left) and
ground truth lesion segmentation (right).  Note that the lesions contained in these
frames increase in size left to right in each row. Lastly, for each row, the first two
image pairs (row 1: case 192; row 2: case 171) correspond to the same lesion observed
from different bronchoscope viewing positions.}
	\label{fig:AFBDatasetOverview}
\end{figure}

In addition, Figure \ref{fig:AFBDatasetOverview} gives sample frames in the training
and validation dataset that illustrate the variations in lesion size ratio.
Each frame pair gives the original frame (left) and ground truth segmentation (right).
As one looks from left to right in the figure, lesion size increases.
We note in passing that the lesion size ratio in the validation dataset is broader
than for the training dataset.  Note that our dataset often draws on
multiple frames depicting the same lesion area.  Yet, these frames show very
different appearances of the lesion, because of inter-frame variations in viewing
direction and distance.  For instance, the first two pairs in Figure
\ref{fig:AFBDatasetOverview} correspond to same lesion appearing in the left lower
lobe of patient case 171:  the second pair is captured from a position closer
to the lesion than the first pair.

\subsection{Polyp segmentation datasets} \label{subsec:Polypdata}

An effective image pattern-recognition model ideally exhibits strong capabilities
for learning and generalizability.  In particular, if a model is trained with a
particular dataset, then the model's learning ability indicates its prediction accuracy.
Similarly, generalizability refers to a model's ability to adapt to previously
unseen datasets.  To ascertain the EPFSNet model's learning capabilities and
generalizability, in Section \ref{sec:results:Polyp}, we
compare our ESFPNet to several state-of-the-art medical image-segmentation models:
Unet++\cite{Zhou-DL2018}, Deeplabv3+\cite{Chen-ECCV2018},
SFA\cite{Fang-MICCAI2019}, CaraNet\cite{Lou-arXiv2021cara},
MSRF-Net\cite{Srivastava-JBHI2022}, and SSFormer\cite{Wang-arXiv2022}. In
particular, for these datasets, we set up three additional experiments summarized below.

\noindent \textbf{Learning ability experiment:} Following the experimental scheme of
MSRF-Net\cite{Srivastava-JBHI2022}, we train, validate, and test ESFPNet on the Kavsir
and CVC-ClinicDB benchmark datasets.  We randomly split each dataset into three
subsets: 80$\%$ training, 10$\%$ validation, and 10$\%$ test.  We freeze the model
when it reaches the optimized dice coefficient on the validation dataset.
We then use the frozen model to generate prediction results for the test dataset.

\noindent \textbf{Generalizability experiment:} We use the same dataset splitting
as recommended in the experimental scheme of ParaNet\cite{Fan-MICCAI2020}, which
used 1450 (90$\%$) video frames from the Kvasir and CVC-ClinicDB datasets for
training.  All images from CVC-ColonDB and ETIS-LaribPolypDB are used for testing.
We keep the best attained performance for each dataset as measures of a model's
forecasting performance on an unseen dataset.

\noindent \textbf{Power balance experiment:} We use the same training dataset as in
the generalizability experiment.  The remaining frames from the Kvasir and CVC-ClinicDB
datasets and all images from CVC-ColonDB, CVC-T, and ETIS-LaribPolypDB are used for
testing.  A model is frozen when it reaches convergence and later utilized in the testing
phase to analyze the segmentation performance over 5 test datasets.

\subsection{Baseline and measurement metrics} \label{subsec:Metrics}

To evaluate our results, we use the following metrics:
mean dice; mean IoU; structural measurement $S_{\alpha}$\cite{Fan-ICCV2017};
enhanced alignment metric $E_{\phi}^{max}$\cite{Fan-arXiv2018}; and the
average mean absolute error (MAE), which gives a measure of pixel-level accuracy.
In addition, we use $S_{\alpha}$ to measure the structural similarity between
predictions and ground truth and the recently proposed  $E_{\phi}^{max}$
to assess both the pixel-level and global-level similarity.  We computed
all metrics using the freely available ParaNet tool \cite{Fan-MICCAI2020}.
When making comparisons to other models, we only use mean dice and mean IoU as the
unifying measures in the learning ability and the power balance experiments.

\section{RESULTS}
\label{sec:results}  

\subsection{AFB dataset experiments}
\label{sec:results:AFB}

We trained ESFPNet-T, ESFPNet-S, and ESFPNet-L on our AFB dataset with batch size 16.
Since the number of lesion and normal frames is not balanced, we added weights on
these two classes to the sampler to guarantee balance when sampling a batch.  We
also trained the Unet++ , SSFormer-S, SSFormer-L, and CARANet models under the
same conditions \cite{Zhou-DL2018,Wang-arXiv2022,Lou-arXiv2021cara}.  Lastly,
results for the R/G method and previously tested machine learning method are given
by Chang {\it et al.} \cite{Chang-EMBC2020}.
Table \ref{table:Comparison} gives the quantitative comparison between various methods.
To help assess a model's complexity and processing time, the table also lists
two attributes of the various models:  number parameters in a model and the amount
of computation (number of floating-point operations) required to derive an output.

Among the various analysis methods, only R/G analysis and our previously proposed
machine learning method missed lesion frames; also these two approaches gave the
largest number of frames falsely identified as lesion frames.  Overall, our proposed
ESFPNet-S model achieved the best segmentation results, while utilizing the third-fewest
network parameters and giving the second fastest execution time.  Although Unet++
requires the second fewest parameters, it generated 5 FP frames.  No other
deep learning network gave any FN or FP frames.  Notably, ESFPNet-S uses the same
backbone as the SSFormer-S.  Yet, it performs better in terms of the mDice and mIoU
indices  while also requiring fewer GFLOPs for computation SSFormer-S.  These
same results also when comparing ESFPNet-L to SSFormer-L.  The CaraNet and
SSFormer-S give nearly identical results, yet CaraNet requires more parameters
and  computation than either SSFormer-S or ESFPNet-S.  Overall, ESFPNet-S
appears to offer the best balance between analysis performance and architectural
efficiency for the AFB dataset.

\begin{table}[!htp]
\begin{center}
\scalebox{0.8}{
\begin{tabular}{c||cllc|cccc}
\multicolumn{1}{l||}{} & \multicolumn{4}{c|}{Architecture Attribute}       & \multicolumn{4}{c}{AFB Results}                                                                                                      \\
Architecture           & \multicolumn{3}{c|}{Parameters} & GFLOPs      & \multicolumn{1}{c|}{Mean Dice}      & \multicolumn{1}{c|}{Mean IoU}       & \multicolumn{1}{l|}{FN Frames} & \multicolumn{1}{l}{FP Frames} \\ \hline
\hline
Masked R/G             & \multicolumn{3}{c|}{N/A}             & N/A          & \multicolumn{1}{c|}{0.549}          & \multicolumn{1}{c|}{0.418}          & \multicolumn{1}{c|}{5}         & 10                             \\ \hline
SVM-EMBC               & \multicolumn{3}{c|}{N/A}             & N/A          & \multicolumn{1}{c|}{0.329}          & \multicolumn{1}{c|}{0.219}          & \multicolumn{1}{c|}{1}         & 26                             \\ \hline
ESFPNet-T (B0)         & \multicolumn{3}{c|}{\textbf{3.5}}            & \textbf{1.4} & \multicolumn{1}{c|}{0.717}          & \multicolumn{1}{c|}{0.574}          & \multicolumn{1}{c|}{0}         & 0                              \\ \hline
ESFPNet-S (B2)         & \multicolumn{3}{c|}{25.0}           & 9.3         & \multicolumn{1}{c|}{\textbf{0.756}} & \multicolumn{1}{c|}{\textbf{0.624}} & \multicolumn{1}{c|}{0}         & 0                              \\ \hline
SSFormer-S (B2)        & \multicolumn{3}{c|}{29.6}           & 20.0        & \multicolumn{1}{c|}{0.746}          & \multicolumn{1}{c|}{0.612}          & \multicolumn{1}{c|}{0}         & 0                              \\ \hline
ESFPNet-L (B4)         & \multicolumn{3}{c|}{61.7}           & 23.9        & \multicolumn{1}{c|}{0.738}          & \multicolumn{1}{c|}{0.600}          & \multicolumn{1}{c|}{0}         & 0                              \\ \hline
SSFormer-L (B4)        & \multicolumn{3}{c|}{66.2}           & 34.6        & \multicolumn{1}{c|}{0.737}          & \multicolumn{1}{c|}{0.604}          & \multicolumn{1}{c|}{0}         & 0                              \\ \hline
Unet++                 & \multicolumn{3}{c|}{9.2}            & 65.7        & \multicolumn{1}{c|}{0.722}          & \multicolumn{1}{c|}{0.587}          & \multicolumn{1}{c|}{0}         & 5                              \\ \hline
CaraNet                & \multicolumn{3}{c|}{46.6}           & 21.8        & \multicolumn{1}{c|}{0.745}          & \multicolumn{1}{c|}{0.610}          & \multicolumn{1}{c|}{0}         & 0                              \\
\end{tabular}
}
\vspace*{+10pt}
\caption[]{\baselineskip=8pt AFB experimental results.
``Parameters" = number of architecture parameters in millions,
GFLOPs = gigaflops (flops = floating-point operations).
The GFLOPs count assumes input dimensions = (1,3,352,352).
FN = frame is misclassified as a normal frame, but is actually a lesion frame
(false negative).  FP = frame is misclassified as a lesion frame, but is actually
a normal frame (false positive). FN and FP results are over 53 lesion and 115 normal
frames, respectively. ``SVM" refers to a machine learning method proposed previously by
Chang {\it et al.} \cite{Chang-EMBC2020}.  ``N/A" = not applicable.
{\bf BOLD} numbers indicate the best measures.}
\label{table:Comparison}
\end{center}
\vspace*{-15pt}
\end{table}

\begin{figure} [!htb]
\vspace*{-12pt}
	\begin{center}
		\begin{tabular}{c}
		\includegraphics[width=6.5in]{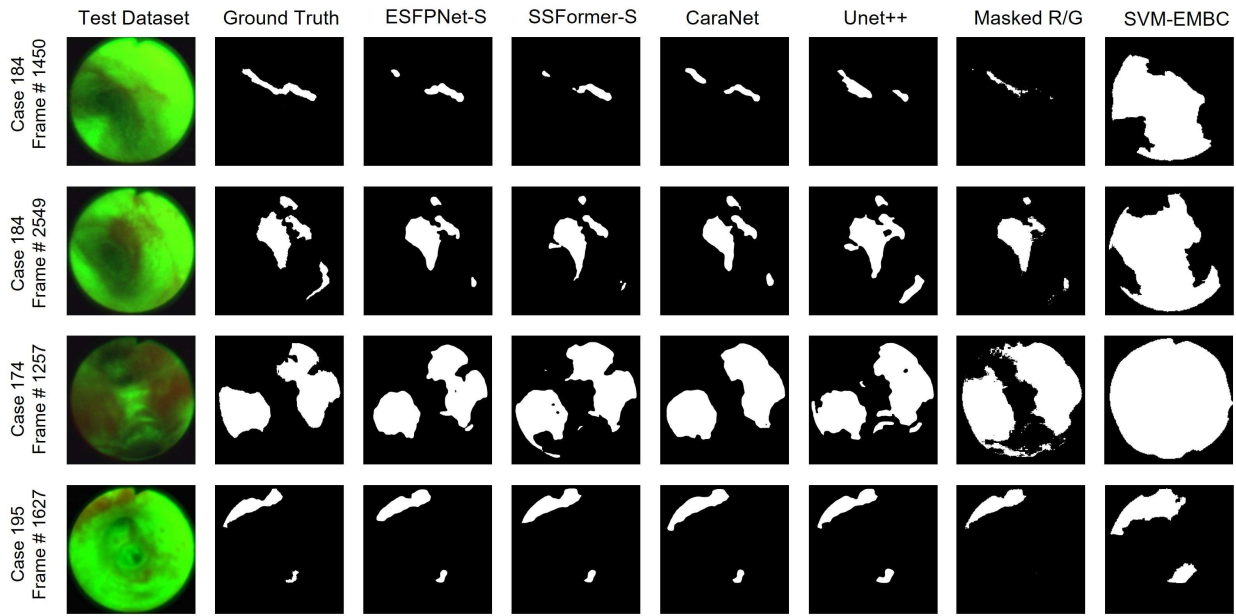} \\
		\end{tabular}
	\end{center}
	\caption[]{\baselineskip=8pt AFB Segmentation results.
The first column is the original frame, and the second column gives the reference
ground truth segmentation.  Results for the other methods are ordered from best to worst.}
	\label{fig:Imageresult}
\end{figure}

Figure \ref{fig:Imageresult} depicts sample AFB segmentation results.  These results
organized in order of the best segmentation results (by ESPFNet-S) to the worst
(by SVM-EMBC); this order also reflects decreasing mDice and mIoU values.  ESFPNet-S gives
the best results among the various methods.  While the masked R/G analysis method,
the clinical baseline standard approach, does detect most lesions, it unfortunately
misses the bottom right lesion in frame 1627 of patient case 195.

\begin{figure} [!htb]
	\begin{center}
		\begin{tabular}{c}
		\includegraphics[width=6.5in]{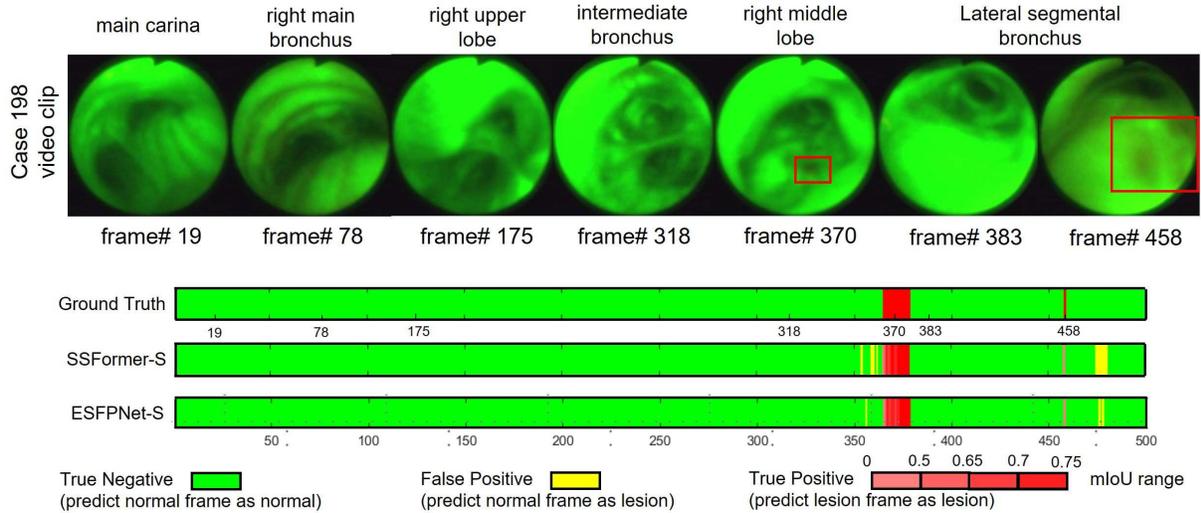} \\
		\end{tabular}
	\end{center}
	\caption[]{\baselineskip=12pt
    Test with a new 500-frame AFB bronchoscopy video clip using SSFormer-S and
    ESFPNet-S.  Both SSFormer-S and ESFPNet-S successfully detect the lesions with high
    IoU values, while ESFPNet-S outputs fewer false positive frames.}
	\label{fig:Videoresult}
\end{figure}

Figure \ref{fig:Videoresult} next illustrates processing results for a 500-frame video clip
from patient case 198, a new case that was not a part of our original train/validate/test
AFB image dataset.  The video clip illustrates a bronchoscopy exam as it starts in the
main carina and then progresses to the right main bronchus and right upper lobe bronchus;
it terminates at the lateral segmental bronchus inside the right middle lobe.  As
highlighted by red boxes in the figure, two lesions have been correctly found in the
right middle lobe (duration $\approx$ 20 frames) and the lateral segmental bronchus
(duration only a few frames).  As the figure clearly shows, both
SSFormer-S and ESFPNet-S successfully detect the lesions with high IoU value, with
ESFPNet-S signaling fewer false positive frames.

As a final test, we implemented the ESFPNet-S into C++ on a Windows-10-based PC,
which includes an Intel Xeon Gold 6230 CPU and NVIDIA RTX 3090 GPU.  With
this set up, we could then test the live real-time performance of the method
for a complete end-to-end segmentation procedure.  The complete procedure consisted
of reading frames from the AFB video stream, image normalization and resizing, ESFPNet
prediction analysis, and displaying the lesion segmentation results on the
computer monitor.  For a 1000-frame test, we achieved an average frame rate of
 27 frames per second (FPS), essentially a real-time rate.

\subsection{Tests on the polyp datasets}
\label{sec:results:Polyp}

Tables \ref{table:LA}-\ref{table:Balance} give experimental results for learning ability,
generalizability, and power balance for the polyp datasets.
These results clearly demonstrate the overall state-of-the-art
capabilities of our proposed ESFPNet-L architecture for polyp segmentation. In particular,
ESFPNet-L  achieved the following:\\
1.) State-of-the-art (SOTA) learning ability on the CVC-ClinicDB benchmark dataset (Table \ref{table:LA}).\\
2.) SOTA generalizability and power balance over the ETIS-LaribPolyDB benchmark data set (Tables \ref{table:GA}-\ref{table:Balance}).\\
Thus, we conclude that our architecture is very much a viable alternative for
polyp segmentation, given its aoverall performance over these various benchmark datasets.

\begin{figure} [!htb]
	\begin{center}
		\begin{tabular}{c}
		\includegraphics[width=5.5in]{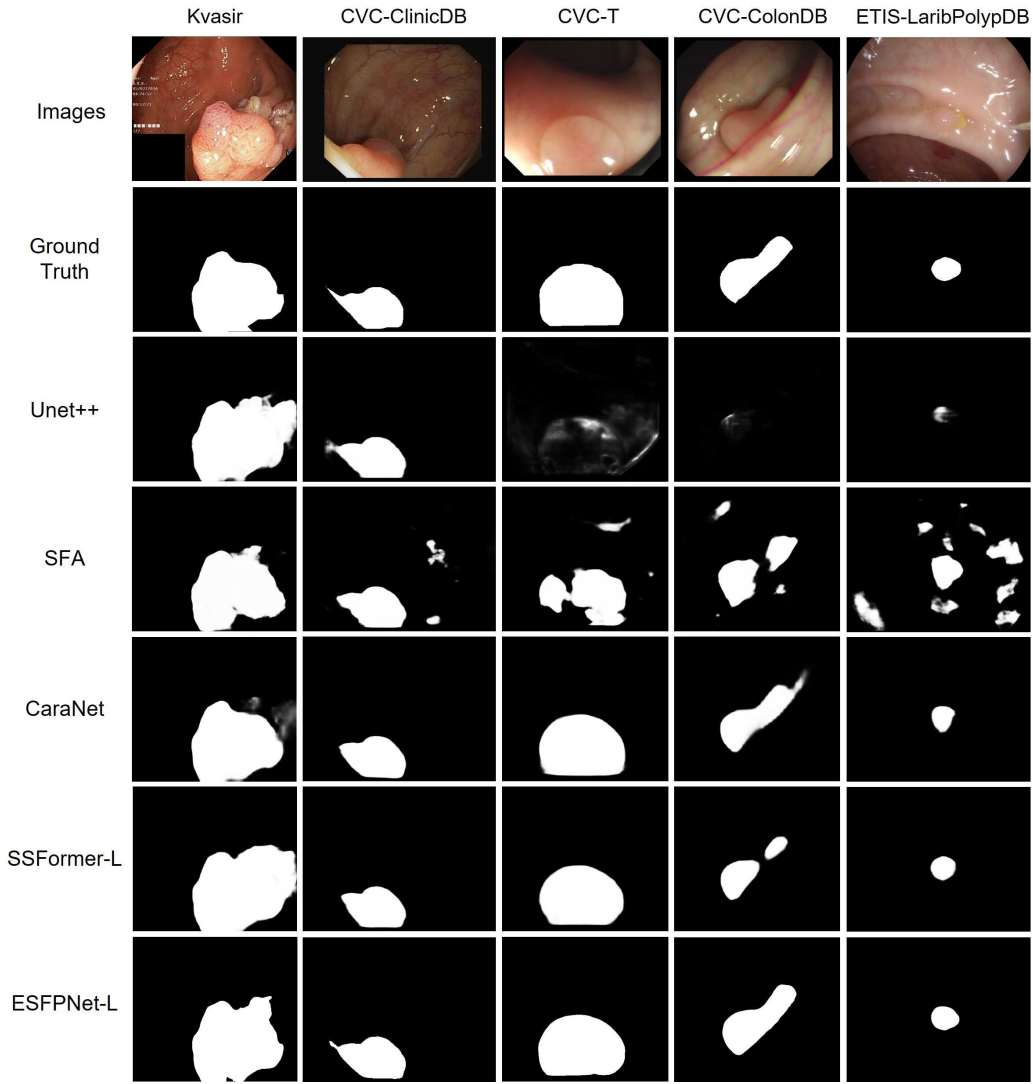} \\
		\end{tabular}
	\end{center}
	\caption[]{\baselineskip=12pt
        Polyp segmentation results.
 }
	\label{fig:Publicresult}
\end{figure}

As final evidence, Figure \ref{fig:Publicresult} gives example polyp segmentation results,
based on the models of Table \ref{table:Balance}.
For the five polyp datasets, our ESFPNet-L shows its outstanding performance on both generalizability and learning ability.

\begin{table}[!htb]
    \begin{minipage}{.45\linewidth}
      \caption{Quantitative comparison of learning ability}
      \label{table:LA}
      \centering
      \vspace{5pt}
      \scalebox{0.8}{
        \begin{tabular}{c||cc|cc}
    & \multicolumn{2}{c|}{CVC-ClinicDB}                    & \multicolumn{2}{c}{Kvasir}        \\
architectures    & \multicolumn{1}{c|}{mDice}          & mIoU           & \multicolumn{1}{c|}{mDice} & mIoU  \\ \hline \hline
U-net++    & \multicolumn{1}{c|}{0.915}          & 0.865          & \multicolumn{1}{c|}{0.863} & 0.818 \\ \hline
Deeplabv3+ & \multicolumn{1}{c|}{0.888}          & 0.871          & \multicolumn{1}{c|}{0.897} & 0.858 \\ \hline
MSRF-Net   & \multicolumn{1}{c|}{0.942}          & 0.904          & \multicolumn{1}{c|}{0.922} & 0.891 \\ \hline
SSFormer-L & \multicolumn{1}{c|}{{\ul 0.945}}          & {\ul0.899 }          & \multicolumn{1}{c|}{\textbf{0.936}} & \textbf{0.891} \\ \hline
ESFPNet-L  & \multicolumn{1}{c|}{\textbf{0.949}} & \textbf{0.907} & \multicolumn{1}{c|}{{\ul 0.931}}      & {\ul 0.887}      \\
        \end{tabular}}
    \end{minipage}%
    \hspace{0.01\linewidth}
    \begin{minipage}{.5\linewidth}
        \caption{Generalizability test.}
        \centering \footnotesize
        \vspace{6pt}
        \label{table:GA}
        \scalebox{0.8}{
            \begin{tabular}{c||c|c|c|c|c|c}
datasets                           & architectures   & mDice          & mIoU           & {$S_\alpha$}           & $E_{\phi}^{max}$         & MAE            \\ \hline \hline
\multirow{3}{*}{\begin{tabular}[c]{@{}c@{}}ETIS-\\ LaribPolypDB\end{tabular}} & ESFPNet-T & 0.781          & 0.701          & 0.866          & 0.910          & 0.016          \\ \cline{2-7}
                                   & ESFPNet-S & 0.807          & 0.730          & 0.879          & 0.916          & 0.015          \\ \cline{2-7}
                                   & ESFPNet-L & \textbf{0.827} & \textbf{0.752} & \textbf{0.892} & \textbf{0.935} & \textbf{0.011} \\ \hline
\multirow{3}{*}{\begin{tabular}[c]{@{}c@{}}CVC-\\ ColonDB\end{tabular}}       & ESFPNet-T & 0.781          & 0.699          & 0.843          & 0.895          & 0.036          \\ \cline{2-7}
                                   & ESFPNet-S & 0.795          & 0.711          & 0.854          & 0.905          & 0.032 \\ \cline{2-7}
                                   & ESFPNet-L & \textbf{0.823} & \textbf{0.741} & \textbf{0.871} & \textbf{0.917} & \textbf{0.029}          \\
            \end{tabular}}
    \end{minipage}
    \footnotesize
    \begin{tablenotes}
        \centering \item {\bf BOLD} values indicate the best results and {\ul underlined} values indicate the second best results.
    \end{tablenotes}
\end{table}

\begin{table}[!htb]
\caption{Power balance experimental results.}
\centering \footnotesize
\label{table:Balance}
\vspace{5pt}
\begin{threeparttable}
\begin{tabular}{c||cc|cc|cc|cc|cc}
\multirow{2}{*}{architectures} & \multicolumn{2}{c|}{Kvasir}                          & \multicolumn{2}{c|}{CVC-ClinicDB}                    & \multicolumn{2}{c|}{CVC-T}                           & \multicolumn{2}{c|}{CVC-ColonDB}                     & \multicolumn{2}{c}{ETIS-LaribPolypDB}               \\
                         & \multicolumn{1}{c|}{mDice}          & mIoU           & \multicolumn{1}{c|}{mDice}          & mIoU           & \multicolumn{1}{c|}{mDice}          & mIoU           & \multicolumn{1}{c|}{mDice}          & mIoU           & \multicolumn{1}{c|}{mDice}          & mIoU           \\ \hline \hline
Unet++                   & \multicolumn{1}{c|}{0.818}          & 0.746          & \multicolumn{1}{c|}{0.823}          & 0.750          & \multicolumn{1}{c|}{0.710}          & 0.627          & \multicolumn{1}{c|}{0.512}          & 0.444          & \multicolumn{1}{c|}{0.398}          & 0.335          \\ \hline
SFA                      & \multicolumn{1}{c|}{0.723}          & 0.611          & \multicolumn{1}{c|}{0.700}          & 0.607          & \multicolumn{1}{c|}{0.297}          & 0.217          & \multicolumn{1}{c|}{0.469}          & 0.347          & \multicolumn{1}{c|}{0.467}          & 0.329          \\ \hline
CaraNet                  & \multicolumn{1}{c|}{\textbf{0.918}} & \textbf{0.865} & \multicolumn{1}{c|}{\textbf{0.936}} & \textbf{0.887} & \multicolumn{1}{c|}{\textbf{0.903}} & \textbf{0.838} & \multicolumn{1}{c|}{0.773}          & 0.689          & \multicolumn{1}{c|}{0.747}          & 0.672          \\ \hline
SSFormer-L               & \multicolumn{1}{c|}{0.917}          & 0.864          & \multicolumn{1}{c|}{0.906}          & 0.855          & \multicolumn{1}{c|}{0.895}          & 0.827          & \multicolumn{1}{c|}{{\ul 0.802}}          & {\ul 0.721} & \multicolumn{1}{c|}{{\ul 0.796}}          & {\ul 0.720}          \\ \hline
ESFPNet-L                & \multicolumn{1}{c|}{{\ul 0.917}}          & {\ul 0.866}          & \multicolumn{1}{c|}{{\ul 0.928}}          & {\ul 0.883}          & \multicolumn{1}{c|}{{\ul 0.902}}          & {\ul 0.836}          & \multicolumn{1}{c|}{\textbf{0.811}} & \textbf{0.730}          & \multicolumn{1}{c|}{\textbf{0.823}} & \textbf{0.748} \\
\end{tabular}
\begin{tablenotes}[para,flushleft]
\centering
{\bf BOLD} values indicate the best results and {\ul underlined} values indicate the second best results.
\end{tablenotes}
\end{threeparttable}
\end{table}

\section{DISCUSSION AND CONCLUSION}
\label{sec:discussion}

Bronchial lesions can be treated as biomarkers that signal lung cancer. These lesions,
which can be detected by autofluorescence bronchoscopy, are useful for detecting lung
cancer at the early stage.  We have proposed ESFPNet for bronchial lesion segmentation
and have applied it to a 20-patient AFB lung cancer patient dataset.  Our results
demonstrate superior segmentation performance over competing deep-learning architectures.
Also, given the ESPFNet's ability to produce results at essentially a real-time frame
rate, it shows potential utility during live bronchoscopic airway exams.  Finally,
to the best of our knowledge, this is the first work for automatic real-time
segmentation of bronchial lesions in AFB video.

As a second result, when ESFPNet was tested for colonic polyp segmentation over five
standard publicly available datasets, it again showed excellent performance.  These
results therefore illustrates its general strong capability for medical image segmentation
in other domains.

The proposed ESPFNet architecture incorporates a novel light-level feature pyramid decoder structure
based on MiT encoders to efficiently and comprehensively use extracted image features.
Through experiments with our AFB dataset, we showed that ESFPNet-S outperforms
other famous models for the mDice and mIoU metrics.  In addition, ESFPNet-S was shown
to be a viable approach for real-time lesion segmentation and detection with AFB video
clips.  Finally, as demonstrated by the colonic polyp segmentation results, ESFPNet exhibits outstanding performance in terms of generalizability and learning ability over
several large datasets.

Nevertheless, further improvements are possible to our work.  First, using only
208 lesion frames to train our deep learning architecture is not realistically
sufficient.  Because of the high cost and difficulty in collecting further live
human video data, we could use semi-supervised learning methods, such as contrastive
learning, to train our model more rigorously.  Second, far more extensive tests on
AFB video clips, in addition to testing the method during a live procedure would
give a more complete understanding of our architecture's overall performance and
viability for live clinical early lung cancer lesion detection.  Third, we need to
create more evidence that our feature pyramid is efficient and comprehensive
in its use of the features extracted by the ESPFNet's encoders; a possibility is to use
attention heat maps to reflect the functionality of each decoding linear prediction
layer.

\acknowledgments 

This work was funded by NIH NCI grant R01-CA151433.
Dr. Higgins and Penn State have financial interests in Broncus Medical, Inc. These financial interests have been reviewed by the University’s Institutional and Individual Conflict of Interest Committees and are currently being managed by the University and reported to the NIH.

{
\vspace{-10pt}
\footnotesize
\baselineskip=6pt
\setlength{\itemsep}{-5mm}
\bibliographystyle{spiebib}
\bibliography{bibtex/Chang,bibtex/mipl}
}

\end{document}